**Graphene-Based Transparent Flexible Strain Gauges with Tunable Sensitivity and Strain Range**

*Joseph Neilson, Pietro Cataldi and Brian Derby*

Department of Materials, University of Manchester, Oxford Road, Manchester M13 9PL, UK

E-mail: brian.derby@manchester.ac.uk

Keywords: strain gauge; reduced graphene oxide; transparent; wearable; flexible

Flexible strain gauges of reduced graphene oxide (rGO) on polydimethylsiloxane membranes, with 88% optical transmittance, are produced from monolayers of graphene oxide assembled into densely packed sheets at an immiscible hexane/water interface and subsequently reduced in HI vapor to increase electrical conductivity. Pre-straining and relaxing the membranes introduces cracks in the rGO film. Subsequent straining opens these cracks inducing piezoresistivity. Reduction for 30 s forms an array of parallel cracks that do not individually span the membrane and a gauge with usable strain range > 0.2 and gauge factor (GF) at low strains ranging from 20 to 100 with pre-strain. The GF reduces with increasing applied strain and asymptotes to about 3, for all pre-strains. Reduction for 60 s leads to the cracks spanning the membrane and an increased film resistance but a highly sensitive strain gauge, with GF ranging from 800 – 16000. However, the usable strain range reduces to < 0.01. A simple equivalent resistor model is proposed to describe the behavior of both gauge types. The gauges show a repeatable and stable response with loading frequencies > 1 kHz and have been used to detect human body strains in a simple e-skin demonstration.

1. Introduction

The ability to measure the elastic or plastic deformation of a given material has a range of applications, including: monitoring response to transient mechanical stress and damage, [1, 2] sensing physiological activity of patients, [3, 4] and e-skins. [5] Traditionally, a strain gauge is fabricated from a metal film and deformation is computed from the variation in electrical resistance that occurs as its length and cross-sectional area changes with strain. The sensitivity of the strain gauge is characterized by its gauge factor, *GF*, which relates the change in electrical resistance, *R*, to the imposed strain, $\varepsilon$, with:

$$GF = \frac{\Delta R}{R_o} \cdot \frac{1}{\varepsilon} \tag{1}$$

Where $R_o$ is the electrical resistance of the unstrained gauge. Conventional metallic strain gauges typically have *GF* values in the range of 1 – 5 and working strain range < 0.05. The principles of operation and applications of the full range of piezoresistive strain sensing has been extensively covered in recent reviews. [6] A number of applications for strain sensing require the sensor structure to be transparent, e.g. structural health monitoring of architectural glass and body mounted motion sensors. There has been some progress in this area using different sensing structures, e.g. carbon nanotube networks with a transmittance of 79% and a GF = 0.4, [7] and multilayer graphene films with a transmittance of 75% and GF = 2.4 up to a strain of 1.8%. [8]

Kang et al demonstrated a novel, highly sensitive strain gauge based on the change in electrical resistance of a Pt thin film deposited on a compliant polymer surface that had been previously elastically strained and relaxed, to generate a population of parallel or channel cracks in the conducting film. It was found that the electrical resistance of this pre-cracked film was very sensitive to subsequent straining, with gauge factors ≈ 1000, with $\varepsilon$ < 0.02. [9] They proposed that the high strain sensitivity is a consequence of the relaxed crack faces coming into partial contact once the cracking strain is removed. Hence, if further deformation occurs, the crack faces begin to separate and there is a period of separation accompanied by a proportional change (reduction) in electrical contact across the crack, which is related to the roughness of the fracture surfaces. This leads to a large sensitivity to strain (large *GF*), until the cracks open sufficiently to break the electrical contact completely. This design of strain gauge has attracted considerable interest in recent years, with various combinations of flexible substrates and conducting films proposed, and this has been extensively reviewed recently. [10] An important distinction between these types of strain sensors is the nature of the cracking patterns that are induced by straining the structure. In the initial report, the channel cracks were observed to extend across the conducting film and multiple cracking leads to a series of approximately parallel and evenly spaced cracks as illustrated schematically in Figure 1a. However, under certain conditions a different cracking scheme is observed with repeated nucleation of cracks that grow and arrest without spanning the specimen, leading to the film being divided into an interconnected network (Figure 1b) which allows a kirigami deformation through the opening of the isolated cracks, without further crack extension. [11]

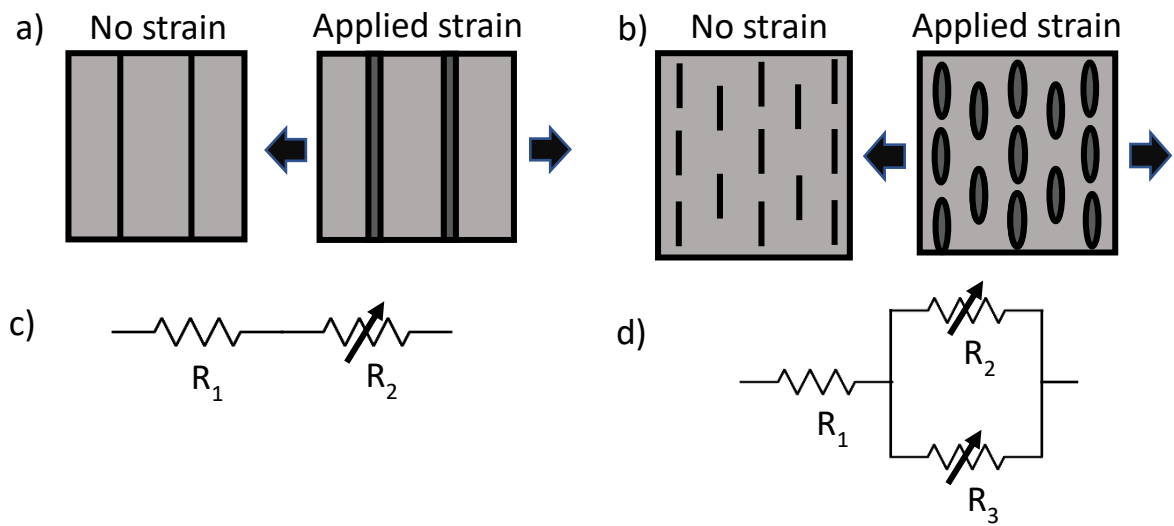

**Figure 1** Schematic representation of crack based strain gauge configurations: a) Parallel cracks spanning the gauge width, b) Kirigami cracking with isolated cracks that do not individually span the gauge, c) Resistor model for parallel cracking, d) Resistor model for kirigami cracking.

The channel crack configuration leads to a high gauge factor (GF > 500) but there is a relatively small strain range before the device is irreversibly damaged ($\varepsilon_{max}$ < 0.02), whereas the kirigami configuration allows much larger extensions before failure ($\varepsilon_{max}$ > 0.2) but with a lower gauge factor (GF < 100). The electrical properties of these crack-based strain gauges can be simulated using appropriate models of resistors in series and parallel as shown in Figure 1. The resistance of the channel crack gauge, $R$, can be adequately modelled by two resistors in series, with a constant low value resistor, $R_1$, representing the resistance of the uncracked film and a much larger variable resistor, $R_2$, representing the resistance across the cracks, which is a function of the crack opening distance and hence the applied strain normal to the crack direction. The resistance of the model gauge is:

$$R = R_1 + R_2 \tag{2}$$

With kirigami cracks there is a current path around each crack as well as one bridging the crack; to account for this, a third resistor, $R_3$, is introduced in parallel to $R_2$, giving the following expression for the resistance of the kirigami cracking configuration:

$$R = R_1 + \frac{R_2 R_3}{R_2 + R_3} \tag{3}$$

A more complex resistor model for *R* was proposed by Jeon *et al.* for kirigami cracking; [11] however, circuit theory can be used to reduce their resistor network to the configuration displayed in figure 1d, which we believe can be used to present a more straightforward interpretation of the piezoresistive response of kirigami cracked films.

Using the crack-based film architecture, it is possible to develop transparent strain gauges through the use of a suitable transparent conducting film. Such transparent, crack based strain gauges have been fabricated using indium tin oxide (ITO), [12] and Ag nanowire networks [13] as the conducting films with an optical transmittance of ≈ 0.9. The ITO film showed a channel crack morphology, leading to a maximum usable strain of ≈ 0.02 and a highly non-linear piezoresistive response with GF increasing from 1 – 1000 over the measured strain range. The Ag nanowire network film showed kirigami cracking and displayed a much larger strain range than the ITO film gauge with GF = 60 at a strain of 1.0. These, crack based architectures, present significantly larger GF values than displayed by the transparent strain gauges fabricated using different piezoresistive mechanisms that were reviewed earlier. [7, 8] Here we present a transparent strain gauge concept based on the generation of cracks in a conductive film of tiled reduced graphene oxide (rGO) flakes deposited on an elastomeric poly dimethylsiloxane (PDMS) substrate, using 2D confined assembly at a planar liquid/liquid interface between two immiscible fluids, as described in earlier work. [14] We demonstrate that through minor changes in the fabrication conditions, the film can be induced to form either a channel crack or a kirigami crack architecture, enabling the formation of both transparent strain gauges of large GF value but small strain range and strain gauges with a large usable strain range but with a lower value of GF.

2. **Materials and Methods**

Graphene oxide (GO) flakes were produced using two-step electrochemical intercalation and oxidation of graphite foil, as described in detail by Cao *et. al.* [15] The GO nanosheets synthesised in this way are predominantly single atomic layers of mean lateral size 3.12 ± 1.27 µm. The as-received dispersion of GO in water was diluted two-fold with isopropyl alcohol (IPA) to generate a 0.05 mg.mL$^{-1}$ GO ink. This was charged to a 10 mL syringe for deposition of tiled monolayers.

PDMS membranes of 500 μm thickness, formed from Sylgard 184 two-part PDMS (Dow, Midland, MI, USA) (polymer to the cross-linker mixing ratio of 10:1 by weight), were deposited on cellulose acetate film (Hartwii, Nanjing, China) by tape-casting with an MSK-AFA-III tape caster (MTI, Richmond, CA, US). After curing at 100 °C in air for 24 hours, the PDMS substrates were treated with a ProCleaner Plus UV-ozone (UVO) plasma cleaner (Bioforce Nanosciences, Salt Lake City, UT, USA) to increase the surface energy of the PDMS surface and facilitate the deposition of GO flakes.

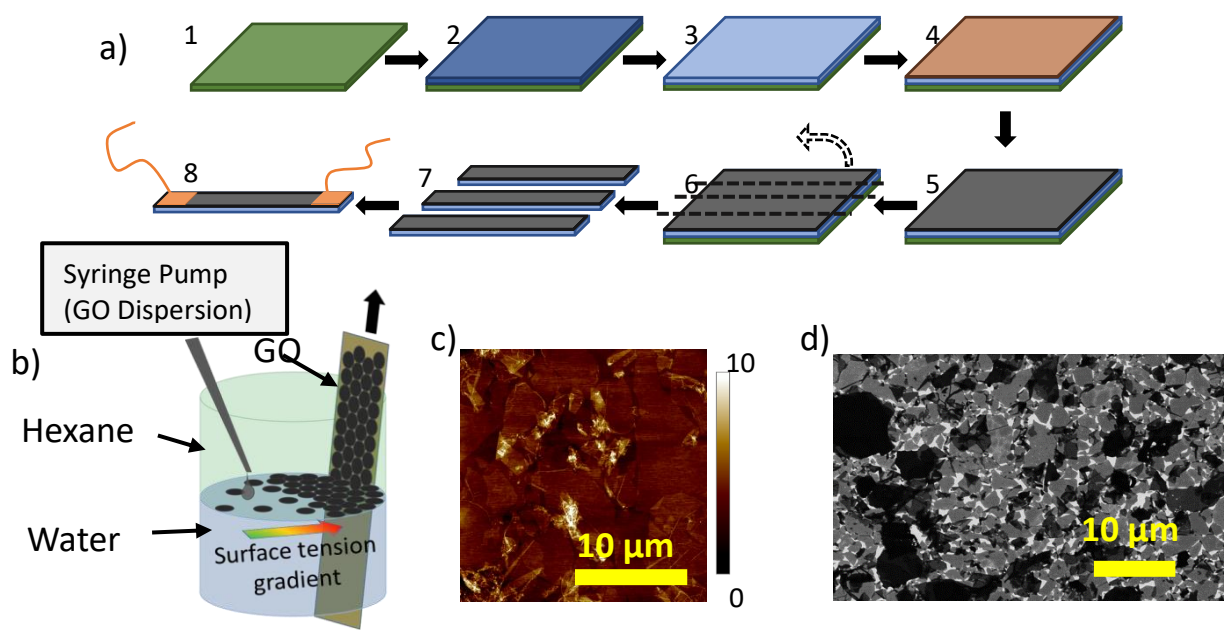

**Figure 2** a) Schematic workflow for the preparation of the transparent strain gauges. b) Assembly at liquid/liquid interface and deposition onto substrate. c) AFM image of the assembled GO monolayer (height range 0 – 10 nm. d) SEM image of the film after reduction to rGO.

Deposition of GO flakes was carried out through assembly at the interface between two immiscible fluids, water and hexane. Previously this method has been used to produce tiled films of MoS$_2$ near-monolayer, 2D material, flakes. [14] Assembly is confined to the 2D boundary of the immiscible interface and a local gradient in surface tension forces the flakes into intimate edge-to-edge contact with minimal overlap between adjacent flakes. This has been described if full detail for the assembly of MoS$_2$ flakes in earlier work. [14] Here we present the essential details of the procedure as used to produce films of GO and highlight where it differs from the procedure used to deposit MoS$_2$. A schematic of the workflow used to produce the sensor structures is presented in Figure 2a.

GO films are formed at the interface between deionised (DI) water, the lower phase, density = 1000 kgm$^{-3}$ and hexane, the upper phase, density = 655 kgm$^{-3}$. A conical flask with a neck internal diameter of 3 cm acted as a reservoir to contain water with the water/air interface about 3 cm below the rim of the flask (Figure S1, Supporting Information). A clamp attached to a precision dip coater positioned the PDMS substrate immediately below the water surface. Hexane is dispensed on the higher density water surface to form a layer of about 5 mm depth. The GO ink in a 50:50 IPA/water suspension was charged into a syringe (10 mL, Sigma) with a 120 mm long needle (21 gauge, Sterican, VWR, Radnor, PA, USA) and inserted into a syringe pump (11 plus, Harvard Apparatus, Holliston, MA, USA). The tip of the needle was positioned at the water/hexane interface. The addition of the GO dispersion was started at 0.1 mL/min and allowed to run for 5 seconds before the PDMS on cellulose acetate was raised by the dip coater at a speed of 1 mm/s. (Figure S2, supporting information) The deposition of the IPA/Water solvent of the ink into the bulk water phase introduces a concentration gradient between the point of injection and the position of the substrate as it is drawn through the liquid/liquid interface (Figure 2b). This gradient in interfacial tension is believed to promote the dense packing of the GO flakes at the liquid/liquid interface and the subsequent transfer of a dense film onto the PDMS substrate. After complete deposition, the GO film was left to dry in a fume cupboard for 1 hour. Figure 2c and 2d show atomic force microscopy (AFM) and scanning electron microscopy (SEM) images of the GO films after deposition and drying. These confirm that the films show minimal overlap of the constituent flakes and that there is continuous edge to edge contact between the flakes providing a continuous path across the film. The dip coating process was used to coat substrates with areas up 25 cm$^2$.

After coating and drying, the GO flakes are reduced to rGO in order to increase the electrical conductivity of the film. This was achieved by placing the GO film on the PDMS substrate into a Schott bottle (500 mL). This was then placed on a hot plate at 70 °C. After 5 minutes to reach thermal equilibrium, 1 or 2 drops of hydroiodic acid solution (≥ 57 % in H$_2$O, Sigma) were dropped into the bottom of the bottle. The bottle was sealed loosely with a lid and the films exposed to HI vapour for times up to 60s. The reduction process converts the film from an insulator to a conductor, with the optical transmittance reducing from 98% to 88% at 550 nm (Figure 3) with a sheet resistance of $R_s$ = 850 ± 42 Ω/□. Sheet resistance measurements were performed using a Jandel 4-point probe system with 1 mm electrode spacing. A source meter (2400 series, Keithley) was used to source and measure current and voltage to the outer and inner probe electrodes, respectively. The optical transmittance of the GO and rGO films were obtained using a Lambda 25 UV-visible spectrophotometer (Perkin Elmer, Waltham, MA, USA).

To fabricate the strain gauges, a 3 cm x 10 cm area of PDMS/cellulose acetate substrate was coated with GO (Figure S2b, Supporting Information). After reduction to rGO, a blade was used to cut individual specimens of dimensions 3 cm x 1 cm. These specimens were carefully removed from the cellulose acetate support and placed in an in-house designed linear extension stage, with displacement resolution of approximately 1 μm. The gauge length of each specimen is defined by the spacing between adhesive conductive copper tape electrodes (3M, Saint Paul, MN, US). The samples were then strained in the linear stage at engineering strains in the range 0.05 - 1.0 to induce cracking in the conductive films. The same stage was also used to measure the change in resistance of the cracked devices after relaxation to zero strain. Crack patterns were imaged and crack spacings measured using an inspection microscope fitted with a CCD camera (AxioCam ERc5s, Carl Zeiss AG, Jena, Germany). A source meter (2400 Series, Keithley, Cleveland, OH, USA) in electrical resistance measurement mode was connected to the copper tape using crocodile clips to measure the devices' electrical resistance change under strain. For the dynamic response measurement of strain gauges, one end of the strain gauge was affixed to the cone of a loudspeaker, and the other end was mounted to a fixed point using adhesive tape, such that the gauges measured the speaker cone displacement in bending mode. The dynamic response was recorded using a Wheatstone bridge circuit connected to an oscilloscope (DSO012A, Agilent Technologies, Santa Clara, CA, USA).

For the devices used in human motion sensing, copper tape was 'cold soldered' to the surface of the devices using silver conductive dispersion (186-3600, RS Pro, Northamptonshire, UK). The gauge was then attached to the reverse of the hand of a subject using adhesive wound dressing tape. The subjects were the authors of the papers, and procedures followed local ethical guidance that required no further approval as no other subjects were used.

3. **Results and Discussion**

After reduction of the GO films with HI vapour, the resulting rGO membrane demonstrates low sheet resistance of $R_S$ = 850 ± 42 Ω/□, and excellent transparency of 88 % at 550 nm (**Error! Reference source not found.**c, d). The optical transparency of the GO is reduced by around 10 % during the reduction process (Figure 2), in line with the findings of other reports, and indicating a restoration of the conducting π-electron system of graphene. [16, 17] Here, the optical absorbance is relatively flat over the entire visible spectrum, suggesting good suitability of the rGO/PDMS membrane as transparent conductive electrodes (TCEs) for applications in large-area optoelectronics devices. Figure 3b compares the sheet resistance and optical transmittance of our rGO/PDMS films with data from previously published work. Top-down indicates films formed by deposition techniques such as spray deposition or spin coating, interface assembly is either at liquid/air (Langmuir-Blodgett) or immiscible liquid interfaces. For full details and sources of the data refer to Table S1, Supporting Information. We believe that the high optical transmittance is the result of good areal coverage with minimal film overlap (Figure 2c and 2d) and the reduced contact resistance through edge-to-edge flake packing. A more detailed analysis of the lack of overlap between 2D material films prepared at immiscible liquid interfaces has been presented in previous work. (14)

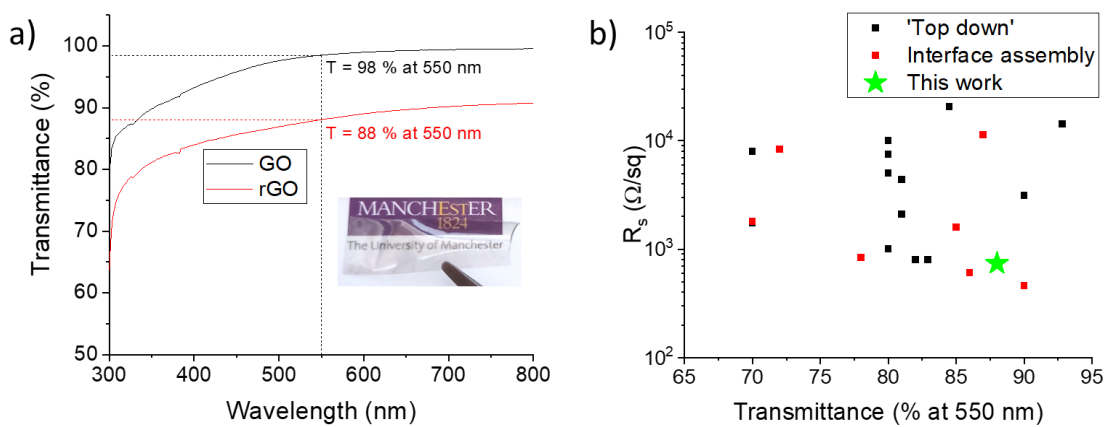

**Figure 3** a) Optical transparency as a function of incident wavelength for the GO/PDMS and rGO/PDMS membranes produced by assembly at the water/hexane interface. b) Comparison of the sheet resistance and optical transparency of transparent, electrically conductive rGO films reported in the literature and this work. See Supporting Information Table S1 for full details and citation information for prior work.

A useful figure of merit for TCEs is the ratio of electrical to optical conductivity,

$$\sigma_{DC}/\sigma_{op} = \frac{Z_0}{2R_s\left(T^{-\frac{1}{2}}-1\right)} \qquad (4)$$

where, $\sigma_{DC}$ and $\sigma_{op}$ are the electrical and optical conductivity of the TCE respectively, $Z_0 = 377\ \Omega$ is the impedance of free space, $R_s$ is the sheet resistance, and $T$ is the optical transmittance. [18] The rGO LAM TCE presented here demonstrates electrical/optical conductivity ratio of $\sigma_{DC}/\sigma_{op}$ = 3.36. This is within the same order of magnitude as some solution processed ITO TCEs, further highlighting the suitability for rGO LAMs in this application. [19] Compared with previous literature examples, the rGO LAMs presented here demonstrate one of the highest reported conductivity ratio (Supporting Information Table S1). However, we note that these rGO LAMs demonstrate the greatest conductivity ratio for rGO films fabricated at maximum processing temperature of less than 100 °C.

### 3.1 Crack Patterns in Strained rGO Films

By initially straining the rGO/PDMS membranes, it is possible to generate the crack structures used to sense strain. [9,10] Figure 4a shows optical micrographs of the films under strain, revealing the morphology of the crack patterns. The UV-ozone treated PDMS membranes show an onset of cracking at a tensile strain $\varepsilon \approx 1$. These cracks nucleate and propagate rapidly across the width of the membrane, normal to the loading direction. As strain increases, further cracks nucleate and propagate to fully span the specimen. This cracking behaviour is commonly called channel cracking. The mean crack spacing, $h$, decreases with increasing strain to a saturation value of approximately 160 μm when $\varepsilon > 1.4$. If the PDMS is exposed to HI vapour for 60 s, cracking initiates at a lower strain and converges to a similar saturation value as found with the as-received PDMS but at lower strain ($\varepsilon > 0.4$). The PDMS/GO and PDMS/rGO membranes show a different behaviour. The PDMS/GO film without HI reduction shows the lowest strain at which cracking initiates. However, in this case the cracks do not propagate across the full width of the specimen but arrest after propagating a few mms normal to the applied load. Further cracks nucleate parallel to the initial cracks and the mean crack spacing decreases to a saturation value of $h \approx 43$ μm at $\varepsilon > 0.2$. The crack pattern is discontinuous and as the applied strain increases the cracks open but do not appear to extend further. We term this behaviour kirigami cracking. After 30 s exposure to HI, the now PDMS/rGO films show a similar behaviour to the PDMS/GO membrane, also forming a kirigami crack pattern. However, the cracks nucleate at a greater initial strain, followed by the mean crack spacing decreasing with increasing strain, in a similar manner to that seen with the PDMS/GO films but over a greater strain interval, reaching saturation at $\varepsilon > 0.5$. After 60 s HI treatment, the PDMS/rGO membrane shows a transition in behaviour, forming channel cracks that span the width of the membrane with the crack spacing saturating close to 160μm.

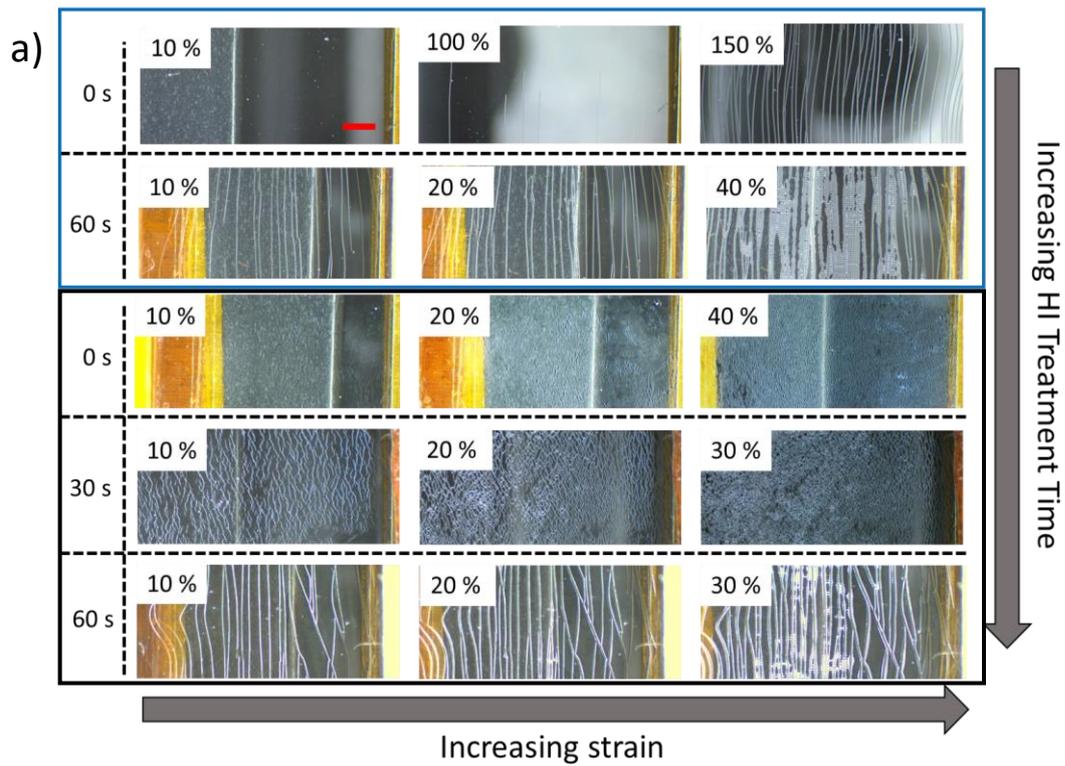

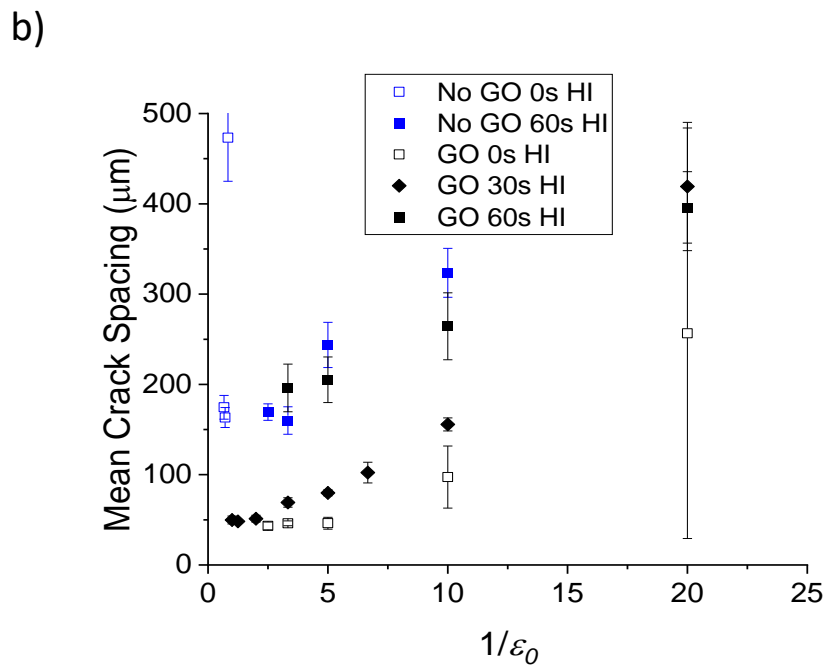

**Figure 4** a) Optical microscopy images of crack morphologies in PDMS, PDMS/GO and PDMS/rGO films as a function of reducing time and applied strain. b) Mean crack spacing after straining as a function of the reciprocal of the applied strain.

The phenomenon of repeated nucleation of channel cracks is well known from studies of the fracture of thin films on compliant surfaces [20] and the solution of Thouless predicts the following relation between strain and mean crack spacing, *h*, for the simplified case when the film and substrate have the same elastic properties: [21]

$$h = 5.6 \sqrt{\frac{t_f \Gamma_f}{\varepsilon^2 \left(E_f/(1-\nu_f)\right)}} \quad (5)$$

where $E_f$, $\nu_f$, $\Gamma_f$ and $t_f$ are Young's modulus, Poisson's ratio, fracture energy and thickness of the film, respectively. Thus, we would expect the crack spacing to be inversely proportional to the maximum applied strain and this behaviour is consistent with our data plotted in Figure 4b. We note that This relationship between crack spacing and the inverse of the applied strain is also found for the kirigami cracked films. However, the analytical solution used to predict this relationship for channel cracking (equation 5) may not be appropriate for kirigami cracking.

### 3.2 Strain Sensing Behaviour

The PDMS/GO membranes show very high electrical resistance and thus are unsuitable for strain sensing applications. The lower electrical resistance PDMS/rGO membranes, formed after exposure to HI for 30 s and 60 s, show extensive cracking patterns and have been tested for their suitability as strain gauges. The different reduction processes have resulted in two distinctly different crack morphologies, which, following the reports on the behaviour of crack based strain gauges in the literature, are expected to show different strain sensing behaviour. [10]

Figure 5a shows the change in resistance of the PDMS/rGO membranes exposed to HI for 60 s as a function of previously applied, or conditioning, strain, $\varepsilon_0$, for a range of conditioning strains up to a maximum of $\varepsilon_0$ = 0.4. In all cases there is a region showing a linear piezoresistive response at low strains with ε < 0.003. In all cases, the strain sensitivity in this regime is approximately constant with GF ≈ 800. The membranes strained to $\varepsilon_0$ = 0.2 and 0.3 both show a sudden upturn in sensitivity at ε ≈ 0.003 to GFs of 16600 and 18000 respectively. Devices at all other conditioning strain levels failed through going open circuit at lower strains (indicated by the coloured arrows in Figure 5a). The resistance of the device after straining and relaxing to zero strain, $R_o$, increases rapidly with increasing $\varepsilon_0$. At $\varepsilon_0$ > 0.3, the devices show much greater initial resistance, with $R_o$ close to the maximum measurable by our equipment, and hence, a reduced sensing range is accessible. A full tabulation of the performance of channel cracked membranes subject to different conditioning strains is presented in Supporting Information Table S2. The GF of these devices at low applied strain is comparable to those reported by others for channel crack strain gauges fabricated from ITO, [12] metal films [9, 22] and Ag nanowires. [23] The transition in sensitivity at high strain is similar to the report of Yang et al. [22] for channel cracked Au/PDMS sensors, who also found a transition to much larger GF values at strains in the range of 0.01 – 0.03. However, they reported that the GF in the low strain regime decreased with increasing conditioning strain.

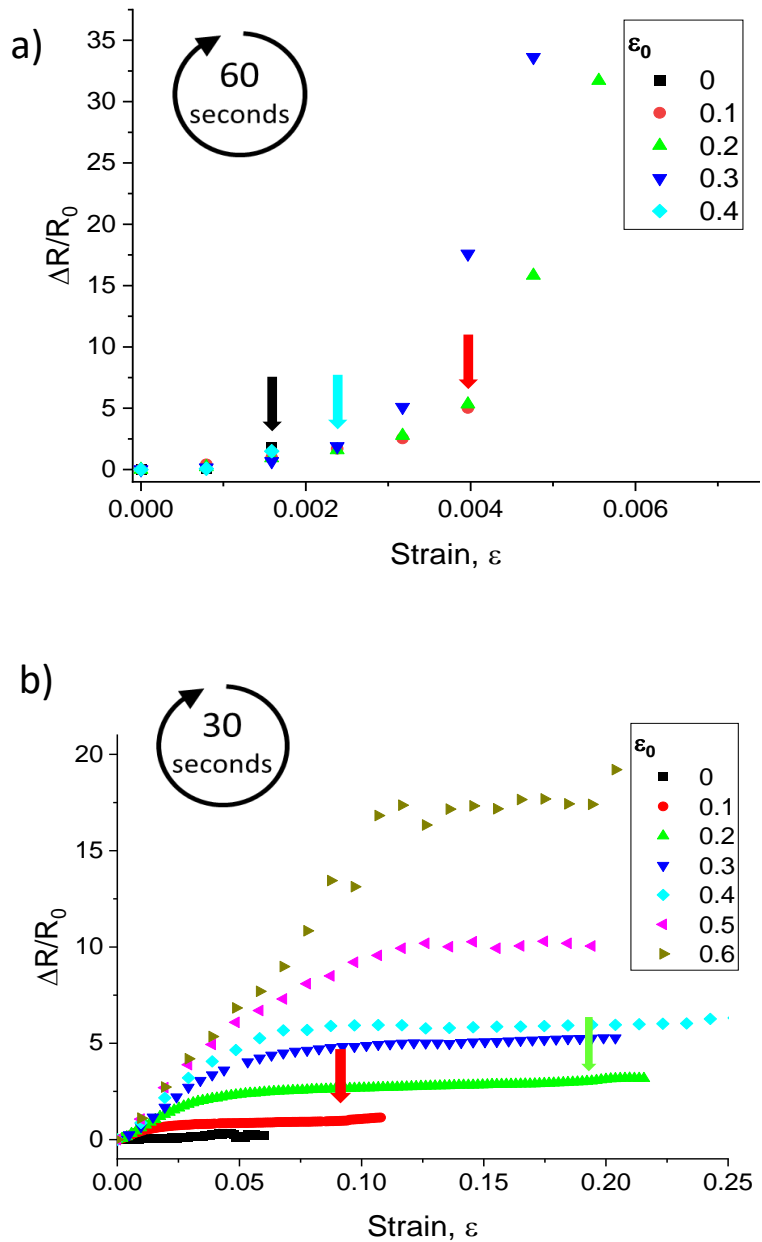

**Figure 5** The change in resistance as a function of applied strain for rGO/PDMS membranes with different amounts of conditioning strain. a) Channel cracking morphology, coloured arrows indicate where gauges went open circuit during testing b) Kirigami cracking morphology, coloured arrows indicate a slight change in gauge factor as the applied strain approaches the conditioning strain.

Figure 5b shows the change in resistance of the PDMS/rGO membranes exposed to HI for 30 s as a function of applied strain, for a range of conditioning strains up to $\varepsilon_0$ = 0.6. These membranes all displayed kirigami cracking. In all case, after applying a conditioning strain the piezoresistive response is highly non-linear, showing a large gauge factor at low strains that gradually decreases with increasing strain to reach a stable value of GF at larger strains. The initial GF increases with increasing conditioning strain, up to a maximum of GF = 149 at $\varepsilon_0$ = 0.6. However, the GF at large strains shows little variation with conditioning strain for $\varepsilon_0$ < 0.6 and, in these cases, the GF ranges between approximately 2 and 4, which is very similar to the GF of the uncracked rGO/PDMS film (Supporting Information, Table S4). The strain range over which the GF decreases to its stable, large strain, value increases with $\varepsilon_0$, but is always smaller than the conditioning strain. With membranes conditioned at $\varepsilon_0$ > 0.6, there is a very small workable strain range and a transition to an irreversible increase in resistance, possibly associated with the formation of channel cracks. Table S3 in the Supporting Information presents a summary of the performance of the kirigami cracked membranes.

The practical working strain range is limited by the conditioning strain, $\varepsilon_0$. If a kirigami cracked membrane is strained $\varepsilon > \varepsilon_0$, there is a noticeable increase in the rate of change of membrane resistance (identified in Figure 5b by coloured arrows). This change is believed to indicate the nucleation of further new cracks in the membrane. We also note that at strains exceeding $\varepsilon$ = 0.2 there is some wrinkling and possible local delamination of the rGO films, this is possibly associated with a mismatch in the elastic properties (Poisson's ratio) of the PDMS and the rGO. The performance of these devices, in terms of sensitivity values (GF), compare well with data from the literature for cracked strain gauges, also based on kirigami crack morphology: e.g. Luo *et al*., using cracked Au films with a carbon nanotube second layer, found a low strain GF = 70 that reduced to GF = 10 at $\varepsilon$ = 0.2 and further still at greater strains; [3] Wang *et al*. used carbon nanotubes as the conducting film but deposited directly onto a PDMS surface before pre-straining to measure a GF = 87 at low strains and GF = 6 at $\varepsilon$ > 0.4. [24] Note that neither of the kirigami network crack sensors in these earlier reports were transparent.

The PDMS/rGO sensors exposed to HI for 30 s and subject to strain conditioning of $\varepsilon_0$ = 0.3 were tested for their repeatability in performance applications at high strain rate and for human motion sensing. Figure 6a shows the repeatable and reproducible piezoresistive response between three of the tested sensors, as indicated by the time-dependant response over three pull-release cycles. The sensors also demonstrate low hysteresis (Figure 6b) and have been shown to work in the strain range of human hand motion (Figure 6c), typically strains in the order of 0.2. [25] The strain sensors also exhibit a fast dynamic response up to 1280 Hz when mounted on an acoustic loudspeaker cone (**Error! Reference source not found.**Figure 6d). This frequency response indicates the possibility to utilise the strain sensors in applications such as human speech monitoring by mounting the devices to the neck. [9] Further, our highly transparent strain sensors would provide an invisible sensing platform for applications in the entertainment and performing arts industries.

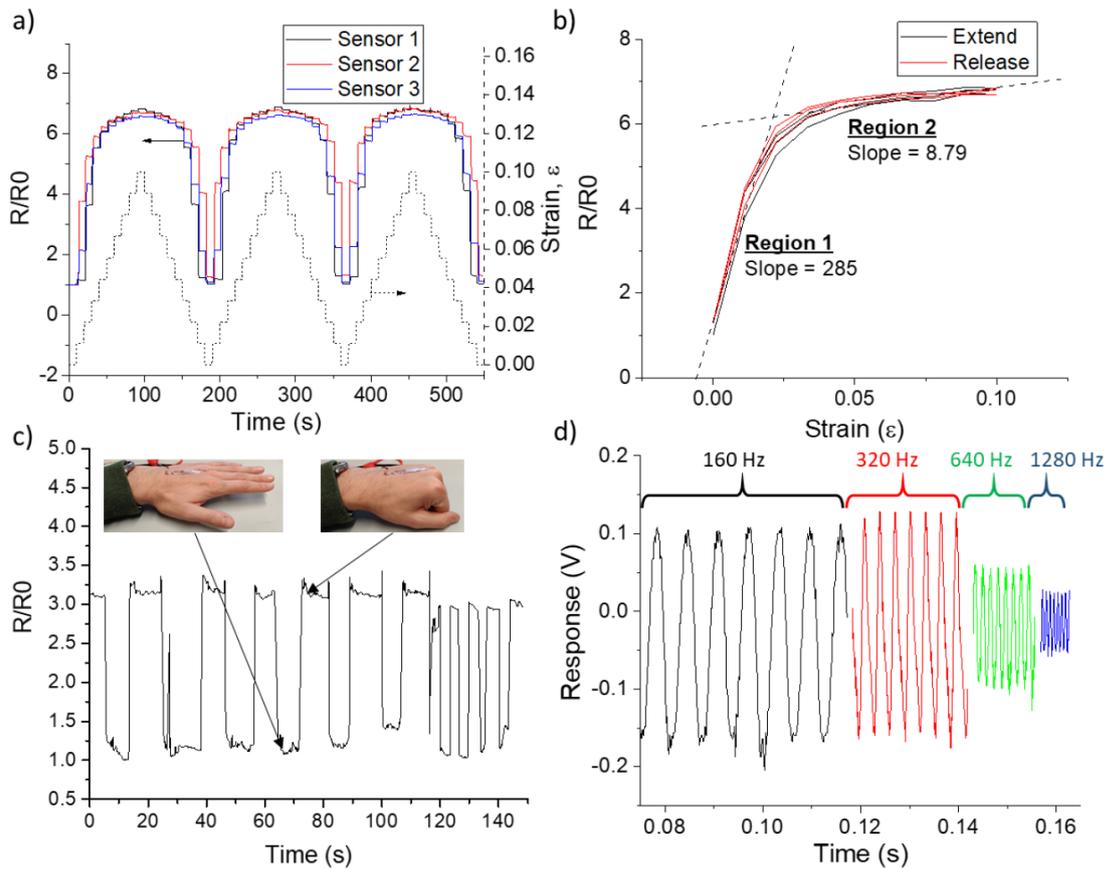

**Figure 6** Strain sensing properties of the 30-seconds HI-treated kirigami-crack-based strain gauges. a) 3x pull release cycles performed with three different strain sensors. The data is displayed as time dependant, and each 'step' is due to applying a 100 um extension to the sensors. The dashed line corresponds to the right-y axis and is the actual strain applied to the sensors. b) Variation of resistance of a sensor extended and released for three cycles. The dashed lines are a result of plotting a linear fit to the two regions of the response curve and the slope correspond to the GF. c) Human motion sensing of the opening/closing of hand by skin-mounted strain sensor. d) Frequency response for a typical strain sensor fixed to a loudspeaker cone.

### 3.3 Mechanisms for the Piezoresistive Effect

*3.3.1 Channel Cracking*

The channel crack morphology has been the most studied in prior work with other conducting films, [9, 10, 12, 22] and this design leads to the highest values of GF but a relatively low workable strain range. The simple series resistor model configuration in Figure 1c presents an appropriate model, if the electrical resistance of a crack is considerably larger than that of the film. Table S2 presents the electrical resistance and mean crack spacing data obtained after straining the PDMS/rGO sensors that were exposed to HI reduction for 60 s, with strains up to 0.3. The resistance per crack, $R^*_2$, is obtained by the following equation:

$$R^*_2 = \frac{R-R_1}{n} \tag{6}$$

where $R_1$ is the resistance of the film prior to straining and $n$ is the number of channel cracks in the membrane. Hence the total resistance of the cracked membrane after the initial conditioning strain, $\varepsilon_0$, is given by the following modified version of equation 2:

$$R_0 = R_1 + nR^*_2 = R_1 + R'_2 \tag{7}$$

Where $R'_2$ is the resistance after relaxation to zero strain contributed by all cracks formed during strain conditioning. From the data obtained from our conditioning strain results (Supporting Information Table S2), the relationship between crack number and conditioning strain is approximately, $n = 190\varepsilon_0$. A simple model for the increase in membrane resistance is that each crack adds a further constant value resistor in series with the contribution from the uncracked film and thus because the number of cracks increases in direct proportion to $\varepsilon_0$, we would expect the membrane resistance to also be proportional to $\varepsilon_0$. However, the data in Table S2 shows that the resistance increases non-linearly with increasing conditioning strain. Thus, the resistance of each channel crack must also independently increase with strain.

Figure 7 shows a logarithmic plot of $R^*_2$, the resistance per crack, as a function of conditioning strain. This suggests that they are related by a power law function with $R^*_2 = K\varepsilon_0^m$, with $m = 5.7$ and a pre-exponent of $K = 10^6$. Combining this function for the crack resistance and using the linear relation between the number of cracks and the initial strain (with $n = h/L_0$ in equation 5), it is possible to use our data and equations 4 – 6 to predict the resistance of the rGO membranes, with parallel channel cracks, as a function of conditioning strain, with

$$R_0 = R_1 + nK\varepsilon_0^{5.7} = R_1 + 1.9 \times 10^8 \varepsilon_0^{5.7} \tag{8}$$

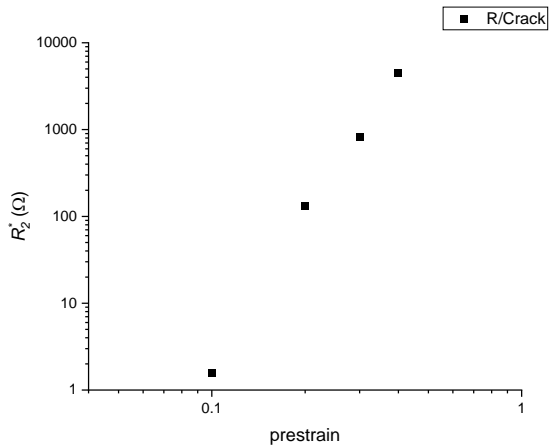

**Figure 7** A plot the mean resistance contributed by an individual channel crack as a function of conditioning pre-strain in the 60s HI treated rGO/PDMS films that show channel cracking.

The mechanism of electrical conduction across a crack in a conducting film has been considered by others and reviewed recently. [10] The mechanisms proposed all rest on the assumption that after the initial conditioning strain, the cracks will close through elastic relaxation when the strain is removed. Thus, the open cracks are replaced by two crack surfaces in contact. Because of details in the film and substrate microstructure, the crack surfaces will not be atomically smooth and thus the nature of the contacting asperities on each surface will lead to local, isolated regions of surface face-to-face contact, regions with no contact, and regions where the two films overlap and are in electrical contact out of plane. These are all captured in the increase in resistance of the cracked membrane once the initial conditioning strain, $\varepsilon_0$, is removed. When the cracked membrane is strained to strains $\varepsilon < \varepsilon_0$ the crack faces will open but no new cracks will be nucleated. The electrical resistance across the crack faces will increase because of a gradual decrease in the electrical contact across the crack faces. [9, 12, 22, 26] An alternative mechanism has been proposed where the conduction across the crack face is driven by electron tunnelling when the crack opening is very small, however this would lead to a highly non-linear piezoresistive response. A transition in mechanism, from asperity contact/film overlap to tunnelling, has been proposed by Luo *et al.* [3] and also by Yang *et al.* to explain the transition to higher GF sensing at larger strains. [22] Hence, we propose that in the channel cracking regime, the series resistor model is consistent with our observations and the resistance of the gauge as a function of strain is described by

$$R(\varepsilon) = R_1 + nR_2^*(\varepsilon) = R_1 + R_2(\varepsilon) \tag{9}$$

Here, the presence of $\varepsilon$ in parentheses refers to the previous parameter being strain dependent and it is also assumed that no new channel cracks are nucleated during straining.

To summarise, when the crack morphology consists of an array of approximately parallel channel cracks in the conducting membrane, the resistance of the membrane under zero load is controlled by the number of cracks normal to the loading direction and the mean electrical resistance of the cracks ($R_2^*$). The number of cracks is determined by the elastic interactions between growing cracks with mean crack spacing proportional to the inverse of the initial conditioning strain, $\varepsilon_0$ (Equation 6). The piezoresistive effect is caused by the change in $R_2^*$ as the cracked membrane is strained. The large increase in membrane resistance with increasing strain is a combination of an increase in the number of cracks and a non-linear relationship between the resistance of a crack and the maximum crack opening that occurs during strain conditioning. The piezoresistive effect at low strains, $\varepsilon < \varepsilon_0$, is the linear response of the individual resistance of the cracks. The low working strain range of these devices is caused by the mechanism for electrical conductivity across the crack faces having a very small linear range and because no new cracks must be nucleated.

*3.3.2 Kirigami Cracking*

The relationship between membrane resistance and conditioning strain is different for the PDMS/rGO membranes reduced for the shorter period of 30 s, which results in kirigami cracking. Note that in this case the cracks are not continuous and the mean crack spacing needs to be carefully defined. This is measured by counting the number of cracks that intersect with a line drawn parallel to the loading direction. It represents the mean distance between a crack and its nearest neighbour in that direction and is not the mean separation of all cracks. Comparing the apparent crack spacing as a function of strain with Thouless's model (equation 5) shows it to be approximately proportional to the inverse of the conditioning strain magnitude (Figure 1b). As with the channel cracked films, the film resistance increases with increasing $\varepsilon_o$ but in comparison, the film resistance remains relatively small (Supporting information Table S3).

The resistance of the kirigami cracked membrane is modelled using a series and parallel resistor model (equation 3). As with the channel crack model, the resistor $R_1$ represents the resistance of the uncracked film, $R_2$ the resistance of the cracks, and $R_3$ the resistance contribution representing the longer current path length caused by the population of discontinuous cracks (Figure 8a). The uncracked film resistance, $R_1$, is constant while the resistors $R_2$ and $R_3$ are assumed to be functions of both the conditioning strain, $\varepsilon_0$, and the sensing strain, $\varepsilon$. Figure 7 shows that the resistance of the channel cracks increases rapidly with strain, and we expect similar behaviour with the individual kirigami cracks. Hence, at large strains, $R_2 \gg R_3$ and Equation 3 reduces to

$$R = R_1 + R_3 \tag{10}$$

Thus at large strains, the asymptotic slope of each gauge in Figure 5b must represent the variation of $R_3$ as a function of strain with

$$R_3 = R'_3 + b\varepsilon \tag{11}$$

Where $R'_3$ is the value of $R_3$ at $\varepsilon = 0$ (the intercept of the asymptotic slope projected to $\varepsilon = 0$) and $b$ is the gradient of the asymptotic slope or the high strain GF. It also follows from equation 3 that the resistance of the membrane at $\varepsilon = 0$ is given by

a)

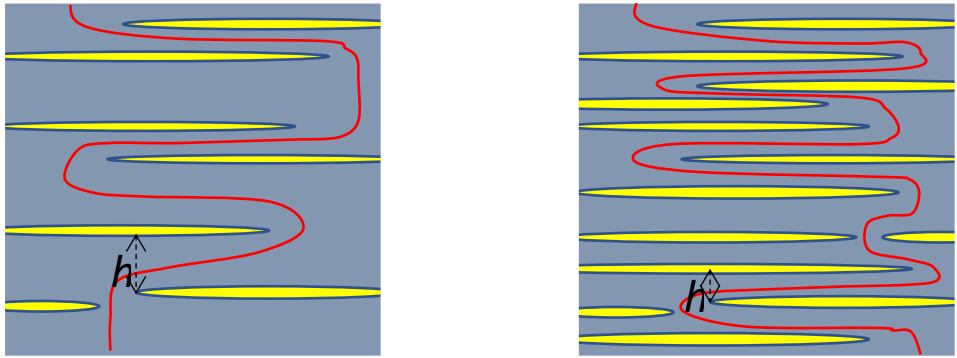

b)

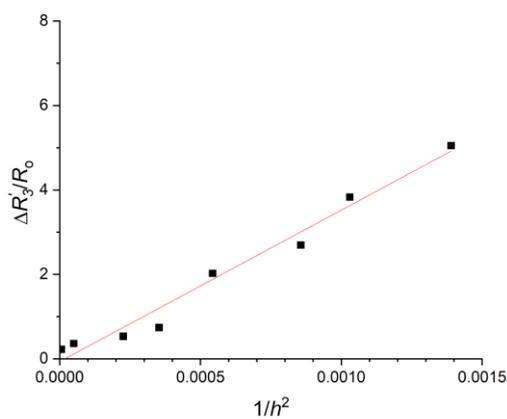

**Figure 8** a) A schematic illustration showing how a reduction in crack spacing ($h$) of kirigami cracks leads to an increase in the current path length within the conducting membrane. b) A plot of the change in $R_3'$ as a function of the square of the inverse of the mean crack spacing in rGO/PDMS films exposed to HI for 30 s and displaying kirigami cracking after strain conditioning.

$$R_0 = R_1 + \frac{R_2' R_3'}{R_2' + R_3'} \tag{12}$$

Where $R'_2$ is the value of $R_2$ at $\varepsilon = 0$. These can be calculated from the data presented in Figure 5 and the appropriate values of $R'_2$, $R'_3$ and $b$ are given in the supporting information Table S3.

Figure 8a shows a schematic of the kirigami cracked membrane at large strains, where $R_2 \gg R_3$. As the crack density increases, the mean crack separation decreases, as does the mean width of the conducting pathway, which is proportional to $h$, the mean crack spacing. If we assume that the mean crack length remains unchanged, the path length for conductivity will increase in proportion to $1/h$. Given that the resistance of the path is proportional to its length and inversely proportional to its width, the increase in resistance, $R_3'$, will be proportional to $1/h^2$, and this is consistent with our measurements (Figure 8b) at conditioning strains, $\varepsilon_0 < 0.8$. Once conditioned, the membrane will contain a distribution of discontinuous, approximately parallel cracks. At zero strain, the cracks will close, and the crack bridging resistance is small. We assume that during subsequent straining as a strain gauge the cracks generated during conditioning do not grow any longer and that no new cracks are nucleated while $\varepsilon < \varepsilon_0$. The resistance of the strain gauge increases from two contributions, a linear increase in $R_3$ as described by the measured gradient $b$ and an increase in the crack bridging resistance that occurs following the kirigami opening of the cracks during straining. This change in resistance is expected to increase rapidly with strain following the behaviour observed with the channel crack devices.

The performance of the strain gauge can be interpreted using a modified form of equation 12

$$R = R_1 + \frac{R_2(\varepsilon) R_3(\varepsilon)}{R_2(\varepsilon) + R_3(\varepsilon)} \tag{13}$$

Where $R_2(\varepsilon)$ is the crack resistance as a function of the applied strain. Figure 9 uses equation 13 to show the response of the gauge and its separation into the contribution of the extra path length, $R_3$, and the crack resistance, $R_2$. This demonstrates that the characteristic form of the strain response curve is shown to be a consequence of the low initial resistance of the closed kirigami cracks ($R_2$) in parallel with resistance caused by the longer path length after kirigami cracking ($R_3$). As strain increases, $R_2$ grows much more rapidly than $R_3$ and thus at high strains the strain response asymptotes to the increase in resistance determined by $R_3$.

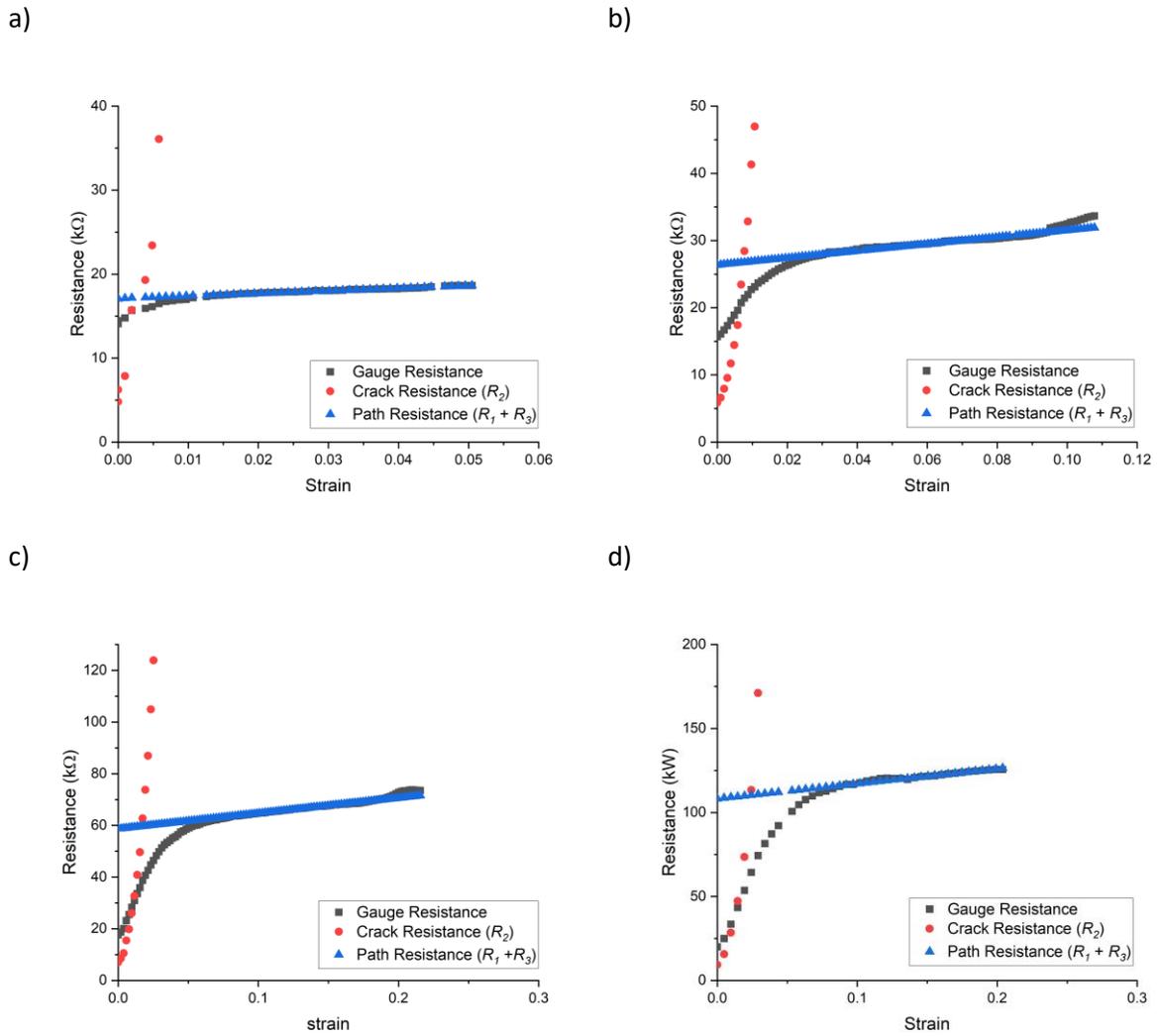

**Figure 9** The strain gauge resistance as a function of applied strain separated into the components determined by resistance across a kirigami crack ($R_2$) and the contribution from the intrinsic gauge resistance and the increased current path length ($R_1 + R_3$), for gauges with conditioning pre-strains ($\varepsilon_o$) of: a) 0.05, b) 0.10, c) 0.20, d) 0.30.

Thus, as with the channel cracking membranes, the strain response is controlled by the very steep increase in the crack electrical resistance as the cracks open during straining. The main difference between the channel cracking and kirigami cracking gauges is that the crack opening resistance is in series with the resistance contribution from the intact membrane when channel cracks form and is in parallel with the membrane resistance when kirigami cracking occurs. Comparing figure 9 with figure 7, it is clear that in both cases the crack resistance increases rapidly with strain but the geometry of the kirigami cracking configuration allows a useful piezoresistive effect to continue to large strains, while that of the channel cracking swiftly rises to an open circuit value, beyond which it is ineffective. The response at large strains for the kirigami cracking gauges is believed to be caused by the gradual reduction in edge to edge contact within the densely packed nature of the randomly tiled 2D monolayer formed during the deposition process. This mechanism is different from the reduction in current carrying area and increase in path length that occurs with conventional strain gauges. The large strain response is possible because of the kirigami nature of the multiple cracks that form, which reduces the likelihood of large channel cracks forming and consequent creation of open circuit conditions. Indeed, in all cases when the kirigami gauges stopped working at high strains, this was accompanied by the nucleation of channel cracks. Hence, a better understanding of the conditions that promote kirigami cracking rather than channel cracking is needed to design gauges with the ability to operate reliably over large strain ranges.

**Conclusions**

We have demonstrated that it is possible to develop a transparent crack-based strain gauge with high gauge factor from a monolayer of graphene oxide flakes, tiled with good edge-to-edge contact and minimal surface overlap, deposited on a thin PDMS elastomeric film. In order to achieve appropriate electrical resistance, the graphene oxide film must be reduced in the presence of HI vapour to form a rGO/PDMS membrane. After reduction the rGO films retain their transparency, allowing 88% of incident light to pass through. The membranes are strained elastically before use as strain gauges to generate a large number of parallel cracks, needed to introduce a piezoresistive response.

The performance of the strain gauges made from these membranes is strongly dependent on the reducing condition. After 30 s reduction by HI, the cracks that form during initial conditioning straining are discontinuous and subsequent straining after relaxation leads to a kirigami opening of the cracks without further crack propagation. However, if the reduction is allowed to proceed for 60 s, pre-straining leads to the formation of channel cracks that spread across the width of the specimen. Both of these cracking morphologies form in rGO films that have high optical transmittance and can be used as transparent strain gauges. The channel cracking configuration results in a very high sensitivity strain gauge with GF > 10,000 but with operating maximum strain < 0.01. The Kirigami cracking gauges show a non-linear but repeatable response up to the initial conditioning strain level when $\varepsilon_o$ is ≤ 0.3. The gauge factor reduces with increasing strain until reaching a value of approximately GF = 3 when the applied strain is about 50% of the initial conditioning pre-strain. With $\varepsilon_o$ > 0.3 there is still a useful operating level but at a reduced strain range. The limiting operating strain for the kirigami devices is either the conditioning strain or the strain at which unwanted channel cracks nucleate and this nucleation is more likely when $\varepsilon_o$ > 0.3.

Human motion sensing is achievable by these sensors due to their Young's modulus (PDMS 1:10, E = 2 MPa) [27] being comparable with that of skin (E = 0.05 - 20 MPa). (28, 29) The strain sensors reported here are ideal candidates for human motion sensing applications in, e.g., e-skins, medical diagnostics, and human-machine interfacing. The PDMS membranes are biocompatible and body conformable, ensuring maximum user comfort, and the high transparency of the devices enables invisible strain sensing. Moreover, the gauge factor and sensing range of the devices can be tuned over almost two orders of magnitude to suit applications where maximum sensitivity is required, such as structural health monitoring.

**Acknowledgements**

We would like to thank Mr Andrew Wallwork for his assistance with fabricating the linear extension stage. We acknowledge the support of EPSRC through grant EP/N010345/1 and for studentship support for JN (ref. 1811873). We also acknowledge the Henry Royce Institute for Advanced Materials, for provision of facilities through EPSRC grants: EP/R00661X/1, EP/S019367/1, EP/P025021/1 and EP/P025498/.

**Graphene based Transparent Flexible Strain Gauges with Tunable Sensitivity and Strain Range**

Joseph Neilson, Pietro Cataldi and Brian Derby

Department of Materials
University of Manchester
Oxford Road
Manchester
M13 9PL, UK

# Supporting Information

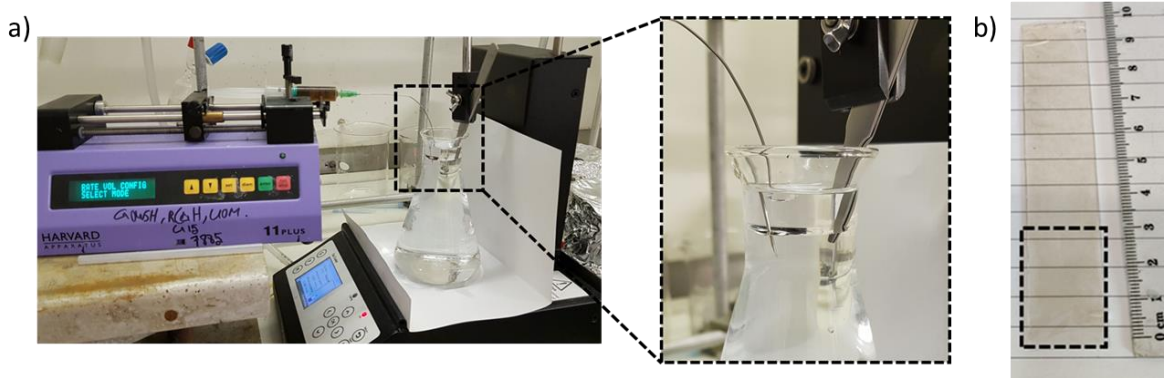

**Figure S1** a) Experimental setup for the continuous production of liquid-liquid assembled monolayers of GO. b) GO film on PDMS. The black box indicates the projected area of the liquid-liquid interface (7.1 cm$^2$) used in the deposition of the GO film of area 25 cm$^2$.

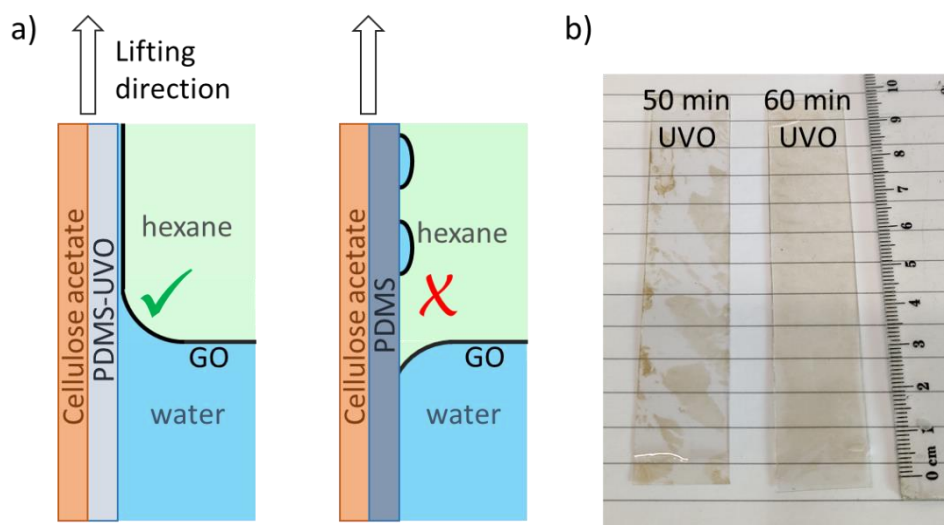

**Figure S2** a) Schematic representation of the continuous deposition of GO via liquid-liquid interface assembly. *Left panel*: the deposition of a continuous GO film relies on a small 3-phase contact angle between the substrate, water, and hexane. *Right panel*: a large 3-phase contact angle will result in beading of the water layer. b) *Left panel*: 50 minute UV-ozone treated PDMS/GO film deposited with a large contact angle showing patchy GO film reduced to rGO. *Right panel*: 60 minute UV-ozone treated PDMS/GO film deposited with a small contact angle showing a uniform GO film reduced to rGO.

**Table S1** Optical transmission, sheet resistance, and corresponding conductivity ratio, $\frac{\sigma_{DC}}{\sigma_{op}}$ for TCE rGO films previously reported in the literature.

| GO Production method | rGO Formation method (Reduction Process) | Film Deposition method | T at 550 nm % | R$_s$ (Ω/□) | $\frac{\sigma_{DC}}{\sigma_{op}}$ | Ref. |
|---|---|---|---|---|---|---|
| Electrochemical | HI Vapour 70 °C 5 mins | Liquid/liquid continuous | **88** | **850** | **3.36** | This work |
| Modified Hummers | Thermal 1100 °C | Spin | 80 | 1000 | 1.59 | 30 |
| Modified Hummers | Hydrazine vapour 400 °C | Spin | 80 | 10000 | 0.159 | 30 |
| Modified Hummers | Chemical: Nascent H$_2$ released by etching of Al with HCl | Spin | 84.5 | 20460 | 0.104 | 31 |
| Hummers | Thermal 1100 °C | Spin | 82 | 800 | 2.25 | 32 |
| Hummers | Thermal 1100 °C under Vacuum | Spin | 80 | 5000 | 0.319 | 33 |
| Hummers | Thermal 1100 °C under Ar and acetylene | Spin | 70 | 1750 | 0.552 | 34 |
| Modified Brodie | NaBH4 solution (150mM) | Spray | 81 | 4400 | 0.386 | 35 |
| Modified Brodie | NaBH4 and AuCl doped | Spray | 81 | 2100 | 0.808 | 35 |
| Modified Hummers | Dispersion containing GO reduced with H2 gas at 50 bar | Spray | 80 | 7500 | 0.213 | 36 |
| Modified Brodie | Roll to roll, SnCl2/EtOH sprayed and 60 °C | Spray | 82.9 | 800 | 2.39 | 37 |
| Ethylbenzoic acid functionalisation of graphite | Thermal 600° C in Ar | Drop cast | 90 | 3110 | 1.12 | 38 |
| Hummers | Thermal 1100 °C under H2 Ar | Dip coating | 70 | 1800 | 0.536 | 39 |
| Hummers | Thermal 1100 °C under Ar | Dip coating | 70 | 8000 | 0.121 | 40 |

| Table S1 continued | | | | | | |
|---|---|---|---|---|---|---|
| GO Production method | rGO Formation method (Reduction Process) | Film Deposition method | T at 550 nm % | $R_s$ ($\Omega/\square$) | $\dfrac{\sigma_{DC}}{\sigma_{op}}$ | Ref. |
| Modified Hummers | HI Solution (55%) 100 °C | Liquid/air (induced by heating) | 78 | 840 | 1.69 | 41 |
| Hummers | Chemical Hydrazine monohydrate 80 °C | Liquid/air (induced by reduction) | 87 | 11300 | 0.231 | 42 |
| Modified Hummers | Immersion in 55% HI 100 °C | Liquid-air (induced by heating) | 85 | 1600 | 1.39 | 43 |
| Modified Hummers with thermal expansion at 1050 °C | Thermally 1100 °C Chemically doped with HNO3 and SOCl2 | Langmuir-Blodgett | 90 | 459 | 7.59 | 44 |
| Modified Hummers | Thermal (1100 C) Under Ar | Langmuir-Blodgett | 86 | 605 | 3.98 | 45 |
| Modified Hummers | Hydrazine 70 °C | Pentane-water assembly | 72 | 8300 | 0.127 | 46 |
| Modified Hummers | Hydrazine (5 %) 95 °C | Toluene-water (induced by reduction) | 70 | 1800 | 0.536 | 47 |
| Electrochemical intercalation and oxidation of graphite foil | HI and acetic acid vapour. | Inkjet printing | 92.8 | 14200 | 0.349 | 15 |

**Table S2**: Electrical resistance and crack spacing of PDMS/rGO membranes reduced for 60 s (channel cracking), measured after applying a conditioning strain, $\varepsilon_0$, and reducing the strain to zero. Specimen gauge length, $L_0$ = 12.6 mm.

| Conding Strain ($\varepsilon_o$) | Mean Crack Spacing ($h$) μm | Crack Number ($n$) | Membrane Resistance ($R$) kΩ | Resistance Increase kΩ | Resistance per Crack ($R^*_2$) Ω | GF Low Strain | GF High Strain | OC Strain | Transition Strain |
|---|---|---|---|---|---|---|---|---|---|
| 0 | | | 9.25 | | | 932 | n/a | 0.0024 | |
| 0.1 | 396 | 32 | 9.30 | 0.05 | 1.6 | 978 | n/a | 0.0048 | |
| 0.2 | 264 | 48 | 15.60 | 6.35 | 133 | 1020 | 16600 | 0.0064 | 0.0048 |
| 0.3 | 205 | 61 | 60.00 | 50.75 | 826 | 1155 | 18000 | 0.0056 | 0.0040 |
| 0.4 | 196 | 64 | 298.00 | 289 | 4491 | 773 | n/a | 0.0024 | |

**Table S3**: Electrical resistance and crack spacing of PDMS/rGO membranes reduced for 30 s (kirigami), measured after different levels of conditioning strain. Specimen gauge length, $L_0$ = 10.3 mm.

| Conditioning Strain ($\varepsilon_o$) | Mean Crack Spacing ($h$) μm | Membrane Resistance ($R_0$) kΩ | Resistance Increase kΩ | GF Low Strain | GF High Strain | $R'_2$ kΩ | $R'_3$ kΩ |
|---|---|---|---|---|---|---|---|
| 0 | - | 11.5 | | 3.7 | | | |
| 0.05 | 419 | 14.1 | 2.6 | 23.4 | 2.2 | 4.8 | 5.6 |
| 0.1 | 156 | 15.7 | 4.2 | 40.3 | 3.3 | 5.8 | 14.8 |
| 0.2 | 80 | 17.6 | 6.1 | 65.0 | 3.4 | 7.0 | 47.3 |
| 0.3 | 69 | 20.0 | 8.5 | 86.8 | 4.5 | 9.4 | 96.9 |
| 0.4 | 60 | 34.8 | 23.3 | 97.2 | 2.1 | 26.1 | 218.4 |
| 0.5 | 51 | 42.5 | 31.0 | 113.3 | 4.0 | 33.4 | 432.5 |
| 0.6 | 50 | 55.6 | 44.1 | 148.8 | 15.7 | 46.4 | 876.7 |
| 0.8 | 48 | 69.6 | 58.1 | - | - | | |
| 1.0 | 50 | 156.9 | 145.4 | - | - | | |

**Table S4**: Literature examples of film based strain gauges showing, strain range, gauge factor and optical transparency, for comparison with our results.

| Material type | strain range | GF (max) | Transparency (%) | Sensing mechanism | ref |
|---|---|---|---|---|---|
| **2D** | **0.006** | **16000** | **88** | **Crack based (transverse)** | **This Work** |
| **2D** | **0.2** | **285** | **88** | **Crack based (kirigami)** | **This Work** |
| 2D | 0.06 | 1000 | NA | Crack based | 48 |
| 2D | 0.071 | 14 | 80 | Crack/defect based | 49 |
| 2D | NA | 4.46 | no | Crack based | 50 |
| 2D | 0.5 | 630 | NA | Changing contact resistance | 51 |
| 2D | 0.004 | 300 | NA | Charge tunnelling | 52 |
| 2D | 0.06 | 66.6 | No | Crack based | 53 |
| 2D | 0.3 | 0.55 | NA | Structural deformation of graphene | 54 |
| 2D | NA | 15 | no | Flake-flake overlap modulation | 55 |
| 2D | 1 | 10 | NA | Percolation modulation | 56 |
| 2D | 0.1 | 9.5 | no | Crack based | 57 |
| 2D | 80 | 20 | no | Percolation modulation | 58 |
| 2D | 800 | 35 | no | Percolation modulation | 59 |
| 2D | 500 | NA | no | Dynamic percolation modulation | 60 |
| 2D | 10 | 2 | NA | Crack based | 61 |
| 2D | 2 | 137 | no | Percolation modulation | 62 |

| Table S4 continued | | | | | |
|---|---|---|---|---|---|
| Material type | strain range | GF (max) | Transparency (%) | Sensing mechanism | Ref |
| 2D | 0.283 | 3.82 | No | Percolation modulation | 63 |
| 2D | 26 | 1054 | No | Crack based | 64 |
| 2D | 550 | 6583 | No | Percolation modulation | 65 |
| 2D | 20 | 42.2 | 89.1 | Crack based | 66 |
| 2D | 25 | 35 | No | Crack based | 67 |
| 2D | 0.5 | 139 | NA | Percolation modulation | 68 |
| 1D | 1 | 4000 | No | Crack based | 23 |
| 1D | 30 | 84.6 | 86.3 | Crack based | 69 |
| 1D | 100 | 30 | 90 | crack based | 13 |
| 1D | 30 | 200 | 92 | Percolation modulation | 70 |
| 1D | 150 | 846 | 88 | Percolation modulation | 71 |
| 1D | 100 | 62.3 | 62 | Percolation modulation | 72 |
| Metal | 2 | 4000 | 89 | Crack based | 73 |
| Metal | 2 | 2000 | No | Crack based | 9 |
| Metal | 2 | 5000 | No | Crack based | 74 |
| Metal | 2 | 1600 | No | Crack based | 75 |
| Metal | 7 | 10000 | No | Crack based | 76 |
| Metal | 2 | 10000 | No | Crack based | 77 |